\documentclass[12pt]{article}
\usepackage{latexsym}
\usepackage{amsmath,amscd}
\usepackage{amsfonts,amssymb}
\usepackage{psfig}
\usepackage{rawfonts}

\newcommand{\nsz}{\normalsize}
\newcommand{\beq}[1]{\large\begin{equation}\label{#1}}
\newcommand{\eeq}{\end{equation}\nsz}
  
\newcommand{\bear}[1]{\large\begin{eqnarray}\label{#1}}
\newcommand{\eear}{\end{eqnarray} \nsz}

\renewcommand{\th}{\hat t} \newcommand{\sh}{\hat s} \newcommand{\uh}{\hat u}
\newcommand{\thmi}{\hat t_{min}} \newcommand{\thma}{\hat t_{max}}
\newcommand{\as}{\alpha_s}

 \newcommand{\JP}{J/\psi}
  
\newcommand{\ac}{\bar c}   
\newcommand{\qq}{q\bar q}  \newcommand{\cc}{c\bar c}

\newcommand{\lit}[3]{{~\bf #2},~#3~(#1)}

\title{%
\rightline{\large{Preprint INP MSU 2001-25/665}}
\vspace*{2.0cm}%
\Large{\bf{A Contribution of Photon Hadronic Component\\
in the Leptoproduction Charmed Structure Function\\
at Large $x$ and $Q^2$}}\\
}

\author{\large
{D.Yu.Golubkov}$^a$\thanks{e--mail: dimgol@mail.desy.de},
\and {Yu.A.Golubkov}$^b$\thanks{e--mail: golubkov@npi.msu.su}
}

\date{%
{\large
$^a$Deutsches Elektronen-Synchrotron (DESY),\\
Hamburg, Germany\\
$^b${Institute of Nuclear Physics of Moscow State University,\\
Moscow, Russia}\\
}%
\vspace*{2.0cm}%
\begin{abstract}%
We calculated the contribution of the photon
hadronic component in the charmed structure 
function of the leptoproduction. The contribution comes from
scattering of the
$c$ quarks of the virtual photon on the quarks and gluons
of the proton. Comparison of our calculations with the measurements
of the charm production in $\mu^+p$-scattering by EMC shows, 
that contribution of resolved photon can explain the excess
of EMC data over the predictions of the 
photon-gluon fusion model at large momentum transfers.
Thus, one does not need to use a non-perturbative admixture
of the charmed quarks in the proton wave function (''intrinsic charm'')
to describe such an excess in EMC data.
\end{abstract}%
}

\begin{document}



\maketitle 

\newpage
\section{Introduction}
\label{introd}

Investigations of heavy flavours production
at high energies remains to provide a very important tool for qualitative and
quantitative test of the QCD and for study of the internal hadron structure.
Despite the impressive experimental and theoretical results
obtained during last decade due to, mainly, experiments at the electron-proton
collider HERA, there still exist uncertainties in the interpretation of
data on the charm production in hadron-hadron and
lepton-hadron collisions. These uncertainties relate to the details
of the quark structure of hadrons, in first turn the proton. One of these
problems is the problem of ''intrinsic charm'' (IC) of the proton
\cite{Brodsky80}, i.e., the question about the presence in the wave function
of the hadron a visible ($\approx\,1\%$) non-perturbative admixture
of the charmed quarks with hard, ''valence-like'' distribution over their
longitudinal momenta. Investigations at HERA have been performed 
at small values of $x$, where the main contribution comes from the photon-gluon
fusion mechanism (PGF). But small values of the variable $x$ describe
the space scales much greater, than the proton size and, therefore,
the study of those kinematical regions gives information about the structure
of the QCD vacuum itself, rather than about the internal structure of the proton.
To understand the internal structure of the proton it is necessary
to study the proton charmed structure function at large values of $x$.
But the geometries of ZEUS and H1 set-ups do not allow
to study efficiently the charm production processes in the very forward 
region at relatively small momentum transfers, where the cross-section
of the $ep$ scattering is maximal. The planned HERA upgrade will increase
few times the collider luminosity and, perhaps, this increase will allow
to collect sufficient statistics at large momentum transfers for the
experimental study of the charm production in the range of large values of $x$.
But at present this problem is not well investigated.

The theoretical consideration of the ''intrinsic charm'' problem is based
on the following simple points.
The wave function of the proton can be expanded over the colourless
eigenstates of free Hamiltonian
$\vert uud>$, $\vert uudg>$, $\vert uudq\bar q>$, $\ldots$.
During sufficiently long time the proton can consist of Fock states
of an arbitrary complexity, including a pair of charmed quarks.
In the proton rest frame the life-time of such a fluctuation $\tau$
has the order of the nuclear time $\sim\,R_h$, where $R_h$ is the hadron size.
Charmed quarks are heavy objects and their life-time is much less than
that of light partons. Therefore, in average, the admixture of the heavy
quark pairs in the proton wave function is to be small,
$\sim\,(m_q/m_Q)^2$, that is, by factor $\sim\,10^2$ less than life-time
of a fluctuation containing light partons only. Additionally, because of
$c$ quarks are massive it turns out that their momentum distributions are
to be much harder than the distribution of the light sea partons 
\cite{Brodsky80,Golub00}. The existence of the fluctuation, containing heavy
quarks is a natural consequence of the field theory. But the important
guess is that quantum fluctuations of the hadron wave function containing 
charmed quarks appear due to self-interaction of the colour field, 
ensuring the non-perturbative origin of the intrinsic charm distribution.
In this case the structure of the hadron Fock states with charm 
can be considered independently of hard interactions.

Estimating the level of the charmed quarks admixture in the proton,
they usually suppose that experimental indications on the intrinsic charm
exist in $\mu p$ collisions \cite{EMC82} and in hadronic collisions 
data \cite{ISRIC,Badier83}, where deviations from the photon-gluon and
parton-parton fusion models predictions have been observed. 
But the experimental situation is sufficiently uncertain. 
The original work \cite{Brodsky80} was motivated by ISR data 
\cite{ISRIC}, where a very large yield of the charm had been obtained, 
at least, order of magnitude larger than it was expected in the model 
of the parton-parton fusion. A critical comparison of the ISR results 
against the contemporary fixed-target data can be found in the review 
\cite{Tavernier87}.
Later experiments (though carried out at lower energies), did not contradict
so dramatically to the theoretical expectations (see, e.g., \cite{Frixione94}). 
In paper \cite{Odorico82} an attempt has been taken 
to co-ordinate the ISR data with the parton-parton fusion predictions. 
There has been considered
the process of the ''charm excitation'', based on the guess
about $0.5\%$ admixture of the charmed quarks in the proton, which admixture
leads to the hard scattering of the charmed quarks.
Due to large cross section of the hard scattering (strictly speaking,
this cross section diverges) and to its stronger dependence on reaction
energy, it was shown \cite{Odorico82}, that ISR data can be described in 
framework of the charm excitation model. At the same time this model 
does not predict visible contribution in the charm yield at lower energies.

Experimental data on longitudinal distributions of charmed particles and
on asymmetry of their production can be reproduced with acceptable accuracy
without the ''intrinsic charm'' hypothesis, only by reasonable variation
of parameters of string hadronization model \cite{Aitala96,Aitala99}.

The interpretation of the data on $\JP$ particle production is also
contradictory. In the experiment NA3 \cite{Badier83} in pion-nuclear
collisions at energy $280$ GeV definite indications on the presence
of an additional mechanism of hard $\JP$ production were obtained.
This additional contribution at the level of $20\%$ of the total of $\JP$ 
production cross section is well described by the modified model of the
intrinsic charm \cite{Golub00}, but the guess about so large contribution
does not look very realistic. It was noted in papers \cite{Vogt95,Biino87},
that intrinsic charm can also explain a larger than it was expected yield
of fast correlated pairs of $\JP$ particles, as well as anomalous polarization
of $\JP$'s, probably observed in NA3 experiment. 
At the same time predictions of the intrinsic charm model directly 
contradict  to data on $\JP$ production in proton-nuclear
collisions at energy $800$ GeV \cite{Kowitt94} (see \cite{Golub96}).

Note, that the admixture of the charmed quarks in the proton
at level of $1\%$ leads to total cross section of charm production in nucleon
collisions of order of few hundreds microbarn in conflict to
the bulk of the experimental data on open charm production in hadronic
interactions.

Doing any theoretical analysis of the hadronic data it is necessary 
to keep in mind, that the main contribution in the charm production in hadronic
collisions is led by the gluon-gluon fusion $gg\,\rightarrow\,\cc$. 
At the same time the direct measurements of the gluon distributions
are impossible. To extract the gluon distributions from data they use
either charm production calculations based on the gluon-gluon and
photon-gluon mechanisms, which contributes in the small $x_F$ range, 
or sufficiently involved methods based on the guess that the main contribution 
in the QCD evolution is driven by the gluons. Due to this fact the extraction
of an additional mechanism of charm production at level of $(0.5-1)$\%
can not be reliable.

In fact, the only experimental data, which allow to perform sufficiently
reliable check of the presence of the intrinsic charm in the proton are
the EMC measurements of the charm photoproduction
in $\mu^+ p$ collisions at energy  $250$ GeV. In early papers on the EMC
data analysis they used old, harder parametrizations of the gluon
distributions in the proton. In particular, in the first analysis carried out
by EMC the scaling gluon distribution $G(x)\,\sim\,(1-x)^5$ has been used, 
which gave the deviation from the photon-gluon mechanism predictions at level
of about $0.3$\% with sufficiently large error of this value.
Modern and, as a rule, softer parametrizations of the gluon distribution,
lead, naturally, to larger contribution of the intrinsic charm in the EMC data
\cite{Golub94,Harris96,Golub00}.
Thus, the intrinsic charm contribution, extracted from analysis
of the EMC data varies from $0.3$\% \cite{EMC82,Hoffmann83} to, approximately,
$0.9$\% \cite{Harris96,Golub00}. In paper \cite{Harris96}, where
the most careful calculations of the contribution of the photon-gluon fusion
have been done, it was shown that at small energy transfers 
the intrinsic charm contribution is absent within the experimental errors.

Due to existence of the hadronic structure of photon 
\cite{Witten77,Bardeen79,Gluck84} the total charm production cross section 
in the leptoproduction should also 
include the interactions of the partons from
the virtual photon with the proton. Such a contribution of the
hard scattering of $c$ quarks from the virtual $\JP$ meson calculated 
in framework of the vector dominance model in \cite{Likhoded00}, allowed to
reproduce with a good accuracy the results of ZEUS experiment on
$D^*$ mesons production at large transverse momenta. Despite the fact that
the scattering processes are to the second order by $\as$,
they can produce quite visible effects due to divergence  
of the parton-parton scattering cross sections at small angles.
Therefore, it is natural to expect, that the hadronic component of the
resolved photon can also give a visible contribution in the charmed
structure function, measured by EMC in the range $Q^2\,\ge\,40$ GeV$^2$, 
where the indications on the deviation of the experimental data from the
photon-gluon predictions exist.

In present work we have calculated the contribution of the
hadronic component of the photon and the contribution of the standard
mechanism of the photon-gluon fusion in the charmed structure function,
measured in EMC experiment. The calculations performed in present work show,
that it is not necessary to consider the intrinsic charm admixture in the
proton to interpret the EMC data.

\section{The basic outline of the model}

\bigskip
\subsection{Charm in the hadronic component of the photon}

In the deeply inelastic processes with hadrons the measured
physical quantity is the hadron structure function $F_2(x,Q^2)$.
The charm production cross section in electroproduction
$Lp\,\rightarrow\,c$, expressed through
the structure function $F_2^c(x,Q^2)$ is:

\beq{eptoc}
d\sigma(l p\,\rightarrow\,c) \ = \ \frac{2\pi\alpha^2}{xQ^4}
\,\left [ 1+(1-y)^2\right]\,F_2^c(x,Q^2)\,dx\,dQ^2\,,
\end{equation}\nsz

\noindent and in the equivalent photon approximation
\cite{Kessler75} it can be expressed through the cross section of the
$\gamma p$ interaction $\sigma(\gamma p\,\rightarrow\,c)$ and the flux
of the equivalent photons $n_{\gamma}(x,Q^2)$ as:

\beq{phflux}
\begin{array}{lll}
d\sigma(l p\,\rightarrow\,c) & = & dn_{\gamma}(x,Q^2)
\,\sigma(\gamma p\,\rightarrow\,c)\,,\\
\\
dn_{\gamma}(x,Q^2) & = & \frac{\alpha}{2\pi}
\,\left [ 1\,+\,\left (1-y\right )^2\right ]\,\frac{dx}{x}\frac{dQ^2}{Q^2}\,.
\end{array}
\end{equation}\nsz

That is, the charmed structure function is connected to the charm
photoproduction cross section as the following:

\beq{pgsci}
F_2^c(x,Q^2) \ = \ \frac{Q^2}{4\pi^2\alpha}
\,\sigma(\gamma p\,\rightarrow\,c)\,.
\end{equation}\nsz

The hadronic structure of the photon can also be described by 
the photon structure function $F_2^{\gamma}(x,Q^2)$, depending on
Bjorken variable $x$ and on the momentum transfer squared $Q^2$ \cite{Witten77}.
The structure function $F_2^{\gamma}(x,Q^2)$ is usually subdivided
in two terms -- perturbative (so called ''anomalous'')
and non-perturbative (''hadronic'') parts. Such a subdivision of  
$F_2^{\gamma}(x,Q^2)$ is valid not only
within the framework of the na\"ive quark parton model (QPM), but also in the 
next-to-leading order of QCD \cite{Witten77,Bardeen79}.
The perturbative part is led by direct interaction
$\gamma\,\rightarrow\,\qq$ and can be completely calculated in the framework
of the perturbation theory. The contribution of this part in QPM is proportional
to $\ln\,Q^2$ and dominates at  $Q^2\,\rightarrow\,\infty$.
The non-perturbative part of $F_2^{\gamma}(x,Q^2)$ is completely analogous to
the ordinary structure function of hadrons. In the framework of the na\"ive QPM
this part does not depend on $Q^2$. The QCD evolution of $F_2^{\gamma}$ 
differs from the evolution of the hadron structure function only due to 
the existence of the perturbative part in the $F_2^{\gamma}$ thus
being described by the inhomogenious DGLAP equations.
To calculate the QCD evolution of the non-perturbative part, similarly to the
hadron case, it is necessary to set the initial conditions at some value
of momentum transfer $Q_0$. As a rule, choosing the initial conditions
for the parton distributions in the photon at
$Q_0^2\,\approx\,1$~GeV$^2$ they suppose, that the hadronic part of
$F_2^{\gamma}(x,Q_0^2)$ can be derived from the vector dominance model (VDM),
postulating the transition $\gamma\,\rightarrow\,\rho^0$ to describe 
the interaction of the photon with hadrons. The parton
distribution in $\rho^0$ meson are obtained using the parton distributions 
in pions known from the experiment. 
For better description of the experimental data
the simple VDM has been extended by the obvious way by inclusion of iso-scalar
vector mesons $\omega ,\,\phi$. The extention of the VMD by inclusion of 
heavier mesons from $\JP$ family is not, normally, used because of large masses
of $\JP$ mesons and the charm content in the photon is described by
the evolution of light partons and by the perturbative term
$\gamma\,\rightarrow\,\cc$ \cite{Gluck84,Schuler93}. But, as it has been shown
in \cite{Gluck84}, in the range $Q^2\,\le\,50$~GeV$^2$ charm must not be
considered as a light quark and the evolution equations must be solved
for three light flavours only. This is because of the fact that in the
threshold region $Q^2\,\sim\,4m_c^2$ the charm production must be described
by the Bethe-Heitler process  $\gamma\,\gamma\,\rightarrow\,\cc$ and by its
analogue, the photon-gluon fusion process $\gamma\,g\,\rightarrow\,\cc$ 
in case of leptoproduction. On the other hand, in paper \cite{Vieira91}
from comparison of the QCD evolution predictions for $F_2^{\gamma}$
to experimental data the conclusion has been made that even for evolution
of light quarks the perturbative region in case of the photon starts at 
$Q_0^2\,\approx\,5$~GeV$^2$, not at $Q_0^2\,\approx\,1$~GeV$^2$.
Thus, in present work we use the generalized
model of vector dominance (GVMD), taking into account also non-perturbative
transitions of the photon to $\JP$ mesons, which become possible beyond
the $\JP$ production threshold.

Then, it is suitable to represent the wave function of the hadronic component 
of the photon as following:

\beq{gamvf}
\vert \gamma_{had} > \ =
\ \sum_V\,\sqrt{\frac{4\pi\alpha}{f_V^2}}\,\vert V>
\,+\,\vert \gamma_{pert}^{(c)}>\,+\,\sum_q\,\vert \gamma_{pert}>\,,
\end{equation}\nsz

\noindent where terms $\vert\gamma_{pert}>$ and $\vert \gamma_{pert}^{(c)}>$
take into account the perturbative contributions of the processes
$\gamma\,\rightarrow\,\qq$ and $\gamma\,\rightarrow\,\cc$, respectively,
and vector mesons $V\,=\,\rho ,\,\omega ,\,\phi ,\,\JP$ describe the
non-perturbative contribution in the hadronic component of the photon.
The constants, characterizing the contributions of vector mesons are
\cite{Schuler93}:

\beq{vconst}
\begin{array}{llllll}
f_{\rho}^2/4\pi & \approx & 2.20\,, 
& f_{\omega}^2/4\pi & \approx & 23.6\,, \\
\\
f_{\phi}^2/4\pi & \approx & 18.4\,, 
& f_{\psi}^2/4\pi & \approx & 11.5\,.
\end{array}
\end{equation}\nsz

At presence of the hadronic structure of photon, the interactions of its
hadronic component can be described in analogue to the 
ordinary hadrons as interactions of the partons from the photon with
the colliding particle. In the framework of the parton model the cross section
of any process is a sum of the cross sections of all elementary subprocesses,
weighted by distribution functions of partons, participating in these
subprocesses. If charmed quarks exist in the proton (''intrinsic charm'')
the leading over strong interaction constant process at moderate momentum 
transfers is the scattering of the lepton on charmed quark
of the proton $lc\,\rightarrow\,lc$. The photon-gluon fusion process
$\gamma\,g\,\rightarrow\,c\bar c$ has the next order in
$\as$, but its contribution dominates due to the fact that gluons carry
almost half of the proton momentum. Thus, one can expect that the contribution 
in the charm photoproduction comes from four processes - the absorption
of the virtual point-like photon by $c$ quark of the proton, 
the photon-gluon fusion, the scattering of the $c$ quark from virtual photon
on light partons of the proton and the scattering of light partons from
the resolved photon on $c$ quarks of the proton. 
Therefore, the total cross section of the charm production can be written as:

\beq{fotcrs}
\sigma (\gamma p)\ = \ \sigma^{IC}\,+\,\sigma^{PGF}
\,+\,\sigma^{\psi}\,+\,\sigma^{(c)}\,+\,\sigma^{(q)}
\end{equation}\nsz

Here, $\sigma^{IC}$ defines the absorption of the virtual photon
by $c$ quark of the proton; the cross section $\sigma^{PGF}$ describes
the photon-gluon process $\gamma g\,\rightarrow\,\cc$. 
The second pair of terms in (\ref{fotcrs}) comes from scattering of $c$ quarks
from the photon on the partons of the proton and $\sigma^{(q)}$ is led by
scattering of light partons from the photon on $c$ quark 
of the proton. The light partons of the photon can originate from 
the non-perturbative part (light vector mesons $\rho ,\,\omega ,\,\phi$), 
as well as from perturbative part $\vert\gamma_{pert}>$.
It follows from (\ref{vconst}), that the contribution of $\JP$ meson
in the hadronic component of the photon is about $16\%$ of the contribution
of the light vector mesons. At the same time the contribution of 
scattering of light partons from photon on charmed quarks from the proton
$\sigma^{(q)}$ is proportional to the level of admixture of the charmed quarks
in the proton, $N_{IC}\,\le\,0.01\,<<\,1$, and can be neglected.
The term $\sigma^{(c)}$ describes the contribution of the scattering of the 
{\em ''perturbative''} $c$ quark from the photon  on partons of the proton.
We denoted as {\em ''perturbative''} those $c$ quarks of the photon,
which appear either due to anomalous part of the photon hadronic component
or to QCD evolution of original light partons.
Thus, the distribution of such $c$ quarks can be obtained in the framework
of the perturbation theory \cite{Gluck84,Aurence94} and, as quoted above, 
this contribution is limited by the range $Q^2\,\ge\,50$~GeV$^2$.
Using the parametrization \cite{Drees85} for the distribution of the
{\em ''perturbative''} $c$ quarks\footnote{Technically, we used the respective
subroutines from the package PYTHIA~6.1 \cite{Pythia61}.}, 
we have verified, that the contribution of the perturbative term $\sigma^{(c)}$
can be safely neglected in the kinematical range considered in present work,
$Q^2\,\le\,80$~GeV$^2$.

\begin{figure}[htb] 

\centerline{\psfig%
{figure=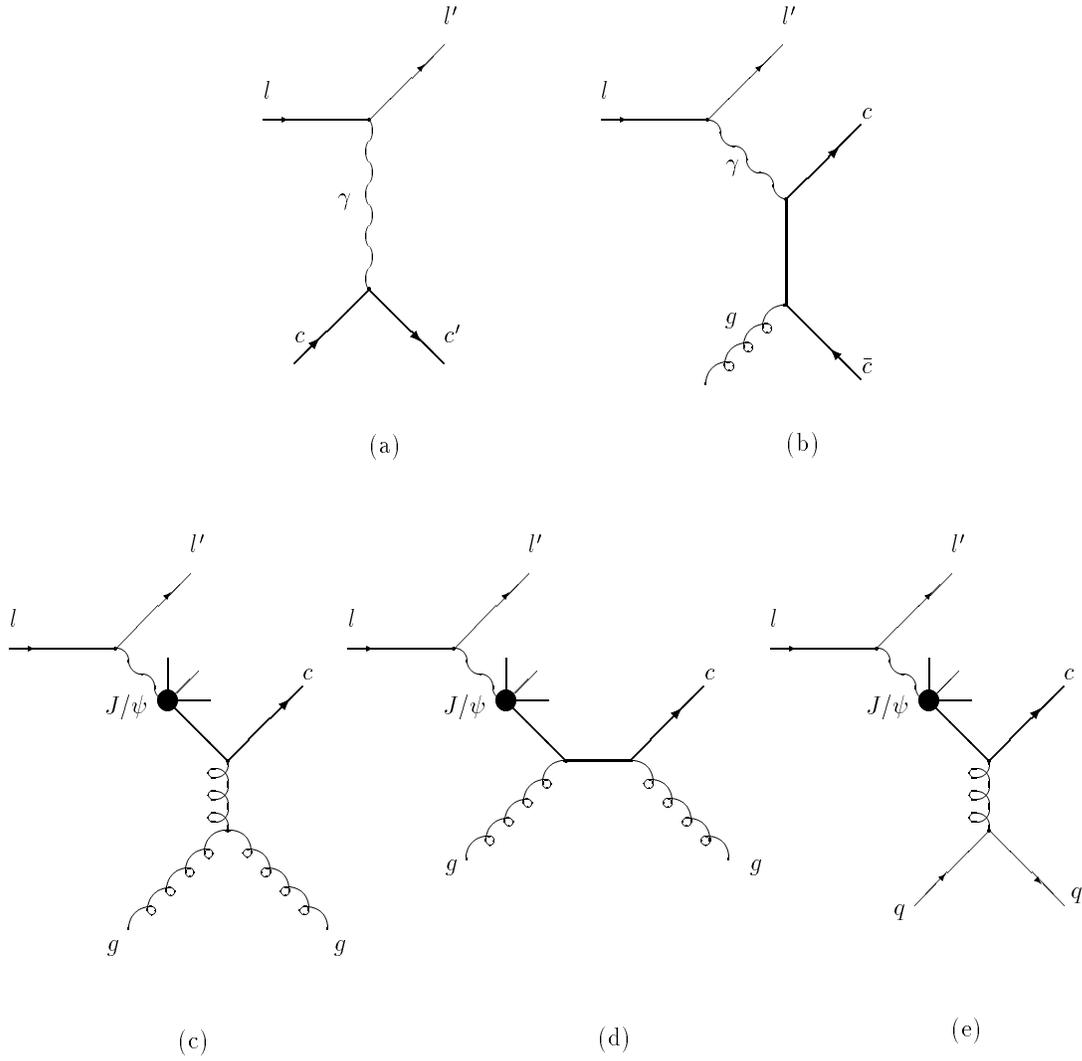,bbllx=0.5cm,bblly=5.0cm,%
bburx=19.5cm,bbury=25.0cm,clip=t,height=16.0cm}
}
\caption{\label{diagrams}
Processes considered in present paper and giving a contribution
in structure function of the charm leptoproduction:
(a) --- the lepton scattering off $c$ quark
(the radiative corrections are not depicted);
(b) --- photon-gluon fusion;
(c), (d), (e) --- the scattering of $c$ quark from the photon on gluon
and quark from the proton, respectively.
}
\end{figure}

The contributions in the charm production come also from the processes
of parton-parton fusion $gg\,\rightarrow\,\cc$ and $\qq\,\rightarrow\,\cc$, 
completely analogous to the processes, describing the charm production
in hadronic collisions. But, these processes have finite total cross
sections and their contribution in the charm yield is small in comparison
to the photon-gluon fusion process. Thus, we also drop them.
Quite different situation appears for scattering processes of $c$ quarks
from $\JP$ meson on light partons of the proton. In the pQCD the cross
sections of these processes are divergent as  $1/p_T^4$.
Therefore their contributions can be significant. As the result of the above
consideration we keep in (\ref{fotcrs}) only terms
$\sigma^{IC},\,\sigma^{PGF}$ and $\sigma^{\psi}$. Their contributions in the
charmed structure function will be calculated further. The diagrams of the
processes, taken into account in present work are depicted in Fig.
\ref{diagrams}.


\bigskip
\subsection{The $c$ quark distribution in virtual $\JP$ meson}

To find the non-perturbative distributions of $c$- and $\ac$ quarks 
in the virtual $\JP$ meson we can follow the approach developed in 
\cite{Golub00}. This approach is based on the statistical model
\cite{Kuti71} and on the work \cite{Brodsky80}, allowing to take into account
the masses of heavy quarks in the framework of the non-covariant perturbation
theory (NPT). This approach allows to find the probability of creation
of the Fock state with heavy quarks. The basic point is
the fact, that we can use NPT to derive the necessary distribution.
Feynman diagram is a sum over all time-ordered diagrams of the NPT.
The contributions of time-disordered diagrams are proportional to the inverse
powers of large hadron momentum $P_h$. Thus, any consideration in the framework
of the parton model is being carried out in the ''infinite momentum 
reference frame'', i.e., in the reference frame, where the hadron momentum 
is much larger than any of characteristic mass parameters.
At high energies of the colliding particles the momentum of the hadron 
is sufficiently large, which ensures the applicability of NPT. 
Only in this case the virtual configuration of Fock states, in which
the proton fluctuates, can be ''frozen'' for the time of the interaction.
If we consider the processes led by the transition of the photon into
a hadron (vector meson), the life-time of the virtual hadronic fluctuation 
(given Fock state) is 
$\Delta t\,\sim\,1/\Delta E\ \approx\ 2\,P_h/\left (M^2-m^2\right )$
($P_h$ - the hadron momentum, $m$ - hadron mass and $M$ - the mass of
the fluctuation) and at high energy can be large enough even for large
values of the fluctuation mass $M$.

Expression (\ref{fotcrs}) means, in fact, 
that different terms in  (\ref{gamvf}) do not interfere.
This, in turn, allows the probabilistic interpretation of the expansion
(\ref{gamvf}). Namely, it is legitimate to assume that during the interaction
time the photon is in one of the states, entering the expansion
(\ref{gamvf}). In this case partons of the photon carry all the momentum
of the photon (more correctly, all the momentum of the hadronic fluctuation).
Due to sharp cut-off over transverse momenta of partons, it is sufficient
to consider the longitudinal phase space only. We consider a hadron
(or hadronic fluctuation of the photon in our case) as a statistical
system, consisting of $m$ partons, carrying in sum the quantum numbers 
of the hadron and its momentum. In spirit of the parton model we suppose
an independent creation of each parton with relative momentum $\xi$
according to the probability density  $\rho (\xi)$. 
Then, in the framework of NPT the probability to find $n$-parton final state
can be written as \cite{Golub00}:

\beq{perprob1}
dW^{(n)}\ \sim\ \frac{1}{\left (E_{fin}\,-\,E_h\right )^2}\,
\delta^{(3)}\left (\vec P_{fin}\,-\,\vec P_h\right )
\prod_{i=1}^n\,\rho_i(\xi_i)\,\frac{d\xi_i}{\xi_i}\,,
\end{equation}\nsz

\noindent where, $n$ is the total number of partons in the considered
fluctuation, $\vec P_h$ and $E_h$ are momentum and energy of the hadron;
$\vec P_{fin}$ and $E_{fin}$ are total momentum and energy of the fluctuation,
respectively.
$\delta$--function enforces the conservation of the total $3$-momentum.
Function $\rho_i(\xi_i)$ describes the probability to create $i$th parton
without conservation of the total momentum of the system and its quantum
numbers. As an argument of the distribution function in the virtual
$\JP$ meson we use the light-cone variable
$\xi_i\,\equiv\,\xi_i^+\,=\,(\varepsilon+p_L)/(E_h+P_h)$, where 
$\varepsilon ,\,p_L$ -- energy and longitudinal momentum of a parton.

Because of the life-time of the fluctuation with a pair of charmed quarks being
much less than the one for the fluctuation, consisting of light partons only,
the full quark-gluon cascade cannot develop. This leads to the situation, 
where the newly created fluctuation with charmed quarks does not practically 
contain light partons and the pair of heavy quarks carries all the
momentum of the virtual $\JP$. Therefore, the probability to observe
the Fock state with $\cc$ pair can be written as follows \cite{Golub00}:

\beq{probhat}
dW^{\cc}\ =\ 
\frac{\xi_c^2\,\xi_{\ac}^2\,\rho_c(\xi_c)\,\rho_{\ac}(\xi_{\ac})}
{\left ( \xi_c\,+\,\xi_{\ac}\right )^2}
\,\delta\left (1-\,\xi_c\,-\,\xi_{\ac}\right )
\,\frac{d\xi_c}{\xi_c}\,\frac{d\xi_{\ac}}{\xi_{\ac}}\,.
\end{equation}\nsz

In $\JP$ meson charmed quarks are the valence quarks. Thus, coming from
Regge phenomenology, one can expect, that at small values of the relative
momentum the probability density $\rho(\xi)$ can be parametrized as:

\beq{rhoval}
\rho_c(\xi) \  = \ \rho_{\ac}(\xi)\ =\ \xi^{\alpha}\,,
\end{equation}\nsz

\noindent where, $\alpha\,\approx\,0.5$, as in ordinary hadrons. So, the
inclusive distribution of the charmed quark in the photon has the form:

\beq{vcrf}
c^{\gamma}(\xi) \ = \ \xi^{\beta}
\ (1-\xi)^{\beta}/B(\beta+1,\beta+1)\,,
\end{equation}\nsz

\noindent with $\beta\,=\,\alpha+1$; $B(u,v)$ is beta-function and
the charmed quarks distributions are normalized on unity,
$\int_0^1\,d\xi\,c^{\gamma}(\xi)\,=\,1$. Choosing 
$\alpha\,=\,0.5$, one obtains $\beta\,=\,1.5$. We shall use this value
in our calculations. The value of $\beta\,=\,1.5$ is smaller than, e.g., 
chosen in \cite{Likhoded00} for description of spectra of charmed mesons,
measured in experiment ZEUS. But in our case the correct form of the 
distribution of $c$ quarks does not play any significant role since
the contribution in the structure function is defined mainly by the value
of the total cross section of the parton scattering. Then, the  
momentum transfers in EMC experiment $Q^2\,\le\,80$ GeV$^2$ practically, are
completely in the range, where the perturbative consideration 
of the production and evolution of $c$ quarks is invalid  \cite{Gluck84}. 
Thus, we do not consider QCD evolution of $c$ quarks in the virtual $\JP$ meson
and we shall use scaling expression (\ref{vcrf}).

\bigskip
\subsection{Charmed structure function in leptoproduction}

Due to the fact that the structure function is up to a common factor
simply sum of cross sections of the subprocesses, it can be also
represented as a sum of contributions of three processes
(see (\ref{fotcrs})):

\beq{fullstrf}
F_2^c(x,Q^2) \ =\ F_2^{IC}(x,Q^2)\,+\,F_2^{PGF}(x,Q^2)
\,+\,F_2^{\psi}(x,Q^2)\,.
\end{equation}\nsz

In framework of parton model the structure function of the ''intrinsic charm'' 
$F_2^{IC}$ is connected to the distribution of charmed quarks in the proton
$c^p(x,Q^2)$ and is equal to:

\beq{f2strf}
F_2^{IC}(x,Q^2)\ =\ 2\,N_{IC}\,e_c^2\,x\,c^p(x,Q^2)\,,
\end{equation}\nsz

\noindent where, $x\,c^p(x,Q^2)$ is normalized on unity, 
$e_c\,=\,\frac{2}{3}$ is the electric charge of $c$ quark and $N_{IC}$ 
defines the level of the admixture of charmed quarks in the proton.

\bigskip
The structure function $F_2^{PGF}$ defines the photon-gluon fusion
and is \cite{Gluck79}:

\beq{f2charm}
F_2^{PGF}(x,Q^2)\ =\ \int_{\sqrt{1+4\lambda}\,x}^1
\,\frac{d\xi}{\xi}\,G(\xi,Q^2)\,f_2(\frac{x}{\xi},Q^2),
\end{equation}\nsz

\noindent where,

\beq{f2gluck}
\begin{array}{lll}
f_2(z,Q^2) & = & \frac{\as(\sh)}{\pi}\,e_c^2\,{\pi}\,z\,
\biggl \{ V_c\,\left [ -\frac{1}{2}\,+\,2z(1-z)(2-\lambda)\right ] \\
& + & \left [ 1-2z(1-z)\,+\,4\lambda z (1-3z)\,-\,8\lambda^2z^2 \right ]
\ln {\frac{1+V_c}{1-V_c}} \biggr \}.\\
\end{array}
\end{equation}\nsz

In the above expressions $\sh\,=\,Q^2\,(1-z)/z$,
$V_c(\sh)\,=\,\sqrt{1-\,4\,m_c^2/\sh}$ is the velocity of 
$c$ quark in the $(\gamma g)$ center of mass system, $\lambda\,=\,m_c^2/Q^2$.

The contribution of the virtual $\JP$ meson, $F_2^{\psi}$, in the charmed
structure function is being led by the expression (\ref{pgsci}) and
the cross section of the process $\JP\,p\,\rightarrow\,c$ in framework
of parton model is:

\beq{gampcrs}
\sigma_{\psi} \ = \ 2\,\,\frac{4\pi\alpha}{f_{\psi}^2}
\sum_i\,\int\,d\xi_i\,d\xi_c\,G^p(\xi_i)
\,c^{\gamma}(\xi_c)\,\hat{\sigma_i}(\hat s)\,.
\end{equation}\nsz

Factor ''2'' takes into account the scattering both on $c$ and $\ac$ quarks.
Index $i$ runs over all partons in the proton, participating in the scattering
off the $c$ quark from the photon. Functions $G^p(\xi_i)$ and $c^{\gamma}(\xi_c)$ 
describe the momentum distributions of the $i$th light parton
in the proton and $c$ quark in the photon, respectively. 
The cross section of the scattering
of $c$ quark on the parton of the proton $\hat{\sigma_i}(\hat s)$ depends on
the CM energy squared $\hat s$. It is known that at presence of
non-perturbative effects, like quark masses, parton model is not a covariant one.
Thus when performing the integration in (\ref{gampcrs}) it is necessary 
to choose some reference frame, satisfying the conditions of the applicabilty
of parton model. In our case the most suitable reference frame is the center
of mass system of the virtual photon and the proton. If we use the light-cone
variable $\xi^+$ the energy squared $\sh$ can be expressed through invariant
variables $x$ and $Q^2$ and is equal to $\sh\,=\,m_c^2+\xi_c\xi_i\,Q^2/x$.

Formulae for the cross sections of the scattering subprocesses
$cg\,\rightarrow\,cg$ and $cq\,\rightarrow\,cq$ of massive $c$ quarks 
in the lowest order over $\as$ can be found in
\cite{Odorico82,Combridge79}. For the completeness we give these formulae
in the Appendix. Pure kinematically the minimal momentum transfer is
$\thmi\,=\,0$. Because the cross section of the scattering process is
divergent at lower limit of the integration over $\th$, the first question
is, obviously, about the cut-off value $\thmi$. At low momentum transfers
pQCD becomes invalid and the value of the cut-off momentum also cannot be
calculated theoretically. For the processes with $c$ quarks 
various choices of $\vert\thmi\vert$ are being used, varying
from $m_c^2 /4$ to $m_{cT}^2$, where
$m_{cT}^2$ is the transverse mass of $c$ quark squared. We shall consider
the value of $\thmi$ as a free parameter, which must be found from experimental
data. We use the same value of $\vert\thmi\vert$ also in the argument
of the strong coupling constant $\as (\vert\thmi\vert)$, because the main
contribution in the total cross section is defined mainly by the range
$\th\,\sim\,\thmi$.

Since we consider the process $\gamma p\,\rightarrow\,\cc\,+\,X$
with large momentum transfer, a coalescence of one of the final $c$ quarks
with the proton remnant is not probable. Thus, according to quantum
numbers conservation, there must be as minimum three particles in the final
state -- the proton and two $D$ mesons. So, general kinematical restriction
for the scattering reaction takes the form ($m_D$ is mass of $D$ meson,
$m_p$ is the proton mass).

\beq{genlim}
\frac{(1-x)}{x}\,Q^2\,+\,m_p^2\ \ge\ (2\,m_D\,+\,m_p)^2\,.
\end{equation}\nsz

At presence of the non-zero cut-off $\thmi$ there are additional kinematic 
constrains on threshold energies $\sqrt{\sh_{th}}$ 
\cite{Odorico82,Combridge79} for the subprocess. 
We also give them in Appendix.

It is easy to obtain necessary integration limits in (\ref{gampcrs}),
considering the kinematics of the parton scattering. Taking into account
non-zero value of $\thmi$ the integration limits over $\xi_i$ and
$\xi_c$ are:

\beq{xgmin}
\begin{array}{lllll}
x\,\left [1\,+\,\frac{(\sqrt{\sh_{th}}+m_c)^2}{Q^2}\right ] &  
\le & \xi_i\ & \le\ & 1\,,\\
\\
\frac{x\,(\sh_{th}-m_c^2)}{\xi_i\,Q^2} & \le & \xi_c & \le & 1\,.
\end{array}
\end{equation}\nsz

\section{Results and conclusion}
\label{concl}

We fitted the calculated $F_2^c(x,Q^2)$ to EMC data. 
The charm photoproduction cross section $\sigma(\gamma p)$,
measured in \cite{EMC82} has been
recalculated to the structure function as described in \cite{Golub00}.
Comparing the calculations with data we took into account 
the next-to-leading order QCD corrections
to the ''intrinsic charm'' structure function $F_2^{IC}$,
as well as non-zero masses
of the proton and charmed quark (see \cite{Golub00,Hoffmann83,Barbi76}).
When calculating the contribution of the photon-gluon fusion we used
the parametrization MRS(G) for the gluon distribution in the proton
from PDFLIB package \cite{PDFlib}. The parametrization MRS(G) has been
obtained with $\as$ in the second order and $\Lambda_{QCD}\,=\,0.174$~GeV.
We included mass of the $c$ quark in free parameters to check 
the self-consistency of our calculations.

\begin{figure}[htb] 
\centerline{\psfig%
{figure=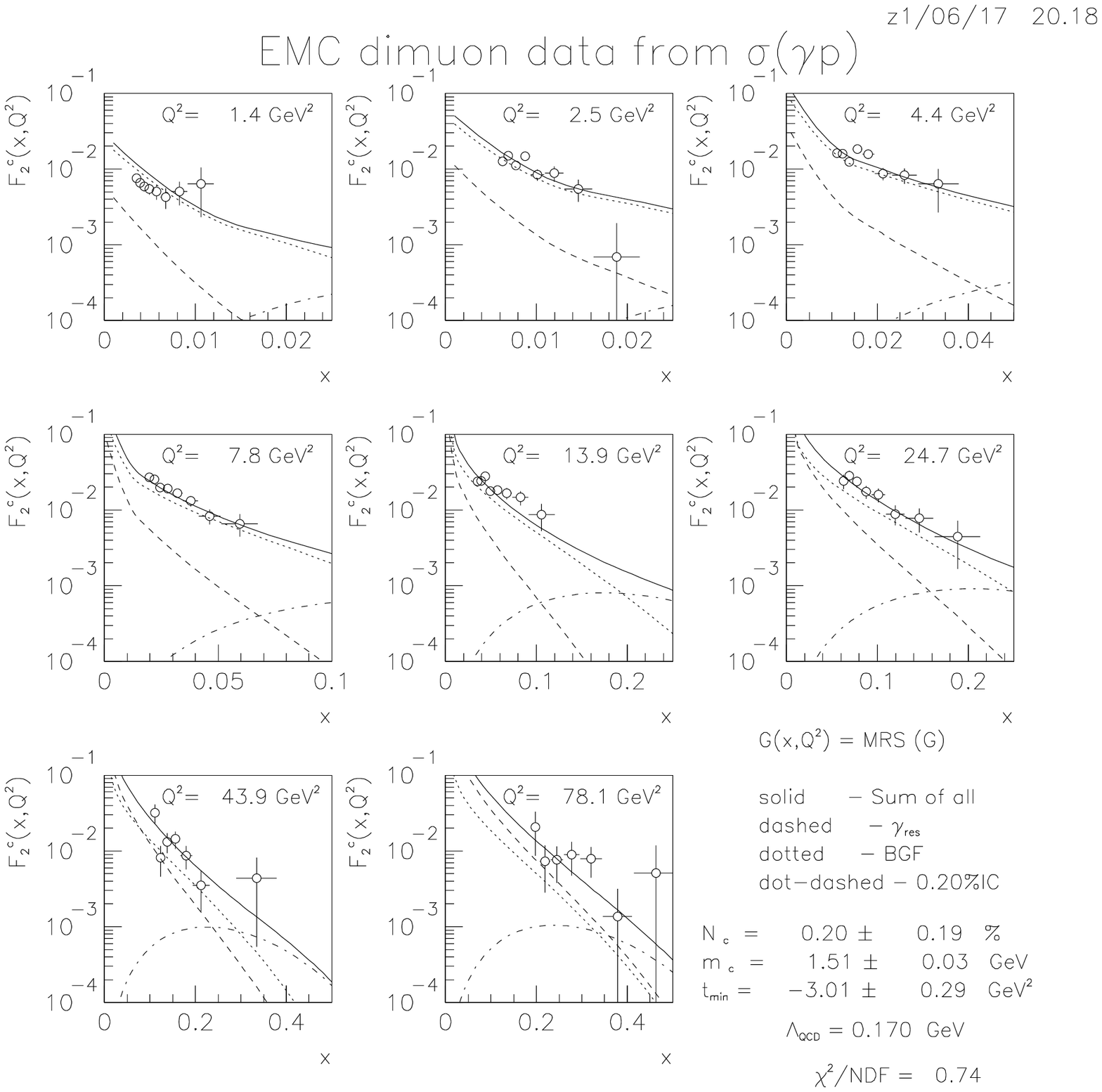,bbllx=0.5cm,bblly=5.0cm,%
bburx=19.0cm,bbury=24.0cm,clip=t,height=16.0cm}
}
\caption{\label{gamplot}
A comparison of the result obtained in present work with experimental
data of EMC. Solid line -- sum of all contributions,
dashed line -- the contribution of the hadronic component of the photon, 
dotted line -- PGF, dot-dashed line -- the scattering of the point-like
photon on $c$ quark from the proton (''intrinsic charm'').
For the gluon distribution the parametrization MRS(G) was used.
The results are given for the following values of the parameters:
$\thmi\,=\,-3.01$~GeV$^2$, $N_{IC}\,=\,0.20\%$ and mass of $c$ quark
$m_c\,=\,1.51$~GeV.
}
\end{figure}

Fit of EMC data gave the following values of free parameters:

\begin{eqnarray}
\label{fitres}
N_{IC} & = & (0.2\,\pm\,0.2)\%\,,\nonumber\\
\\
m_c & = & (1.51\,\pm\,0.03)\ \mbox{GeV}\,,\nonumber\\
\\
\thmi & = & (-3.0\,\pm\,0.3)\ GeV^2\,,\nonumber
\end{eqnarray}

\noindent at good value of $\chi^2/NDF\,=\,0.74$.
The value of cut-off over the momentum transfer squared is approximately equal
to the transverse $c$ quark mass squared, as it could be expected from
general considerations. The resulting value $m_c\,=\,1.51\,\pm\,0.03$~GeV 
agrees with $m_c\,=\,1.50$~GeV, used in the parametrization {\rm MRS(G)}, 
which testifies the self-consistency of our approach. The comparison
of the model with EMC data \cite{EMC82} is given in Fig.\ref{gamplot}.

Note the following. In paper \cite{Odorico82} the calculations have been
performed for the charm excitation mechanism in hadron collisions, i.e.,
there have been considered the processes of the scattering of partons
on the charmed quarks of the proton (on ''intrinsic charmed quarks''). 
The contribution in the charm leptoproduction formed by 
hadronic component of the photon is completely analogous 
to the charm excitation in hadron collisions,
considered in \cite{Odorico82}. In paper \cite{Odorico82} the level of the
intrinsic charm in the proton has been chosen equal to $0.5$\% and the
conclusion was made that the charm excitation mechanism ensures
the needed yield of charmed particles at ISR energies due to steeper rise
of the scattering cross section with reaction energy, which allows
to agree ISR data and data obtained at lower energies in fixed target
experiments. In paper \cite{Odorico82} the cut-off momentum transfer has been
chosen equal to $\thmi\,=\,m_c^2/4\,\approx\,0.6$ GeV$^2$, which is
considerably less than obtained in present paper.

The results of present paper give significantly less admixture of
the intrinsic charm in the proton and much larger value of the cut-off momentum 
$\thmi$ in the scattering subprocesses. Therefore, the contribution of the
charm excitation mechanism in the hadron collisions is, as minimum, order of
magnitude less than that obtained in \cite{Odorico82} and cannot explain
large yield of charmed particles, observed in ISR experiments.
Thus, the contradiction between ISR data and data of fixed target experiments
is not avoided.

As it follows from the results of present work, it is not necessary
to introduce any visible admixture of the non-perturbative
charm quarks in the proton to describe EMC data, even when using ''soft''
gluon distribution. The excess of the EMC data over
the photon-gluon predictions at large $Q^2$ can be completely explained by
the contribution of the hadronic component of the photon. In any case,
the considered mechanism decreases significantly the discrepancy between
EMC data and PGF predictions and, respectively, the admixture
of Fock states with charmed quarks in the proton.

On the other hand, we have introduced a non-perturbative charmed component
in the photon wave function. The reason, why the photon can have visible
non-perturbative component of $c$ quarks  in difference to the proton
can be understood from result of paper \cite{Vieira91}, which states that the
perturbative regime in the photon starts at $Q^2\,\approx\,5.5$~GeV$^2$,
and for the proton the DGLAP equations give good description of the evolution
of the structure function starting already from  
$Q^2\,\approx\,(1\,-\,2)$~GeV$^2$. Thus, there is no necessity to introduce
the non-perturbative charmed states in the proton and the charm contribution
to the proton structure function can be described by pQCD.

\bigskip 
\vbox{
\centerline{\Large Acknowledgement}

\bigskip
We would like to express our gratitude to DESY for hospitality
and good working conditions.
}%

\vfill\eject



\newpage
\section{Appendix}

Expressions for total and differential cross sections of $c$ quark 
scattering at lowest order of $\as$ \cite{Combridge79}
(we write out these formulae as they are given in \cite{Odorico82}).

\large

$$
\begin{array}{lll}
\frac{d\sigma(\qq\,\rightarrow\,\cc)}{d\th} & = & 
\frac{4\pi}{9\sh^2}\,\as^2(4m_c^2)
\,\frac{(m_c^2-\th)^2+(m_c^2-\uh)^2+2m_c^2\sh)}{\sh^2}\,;\\
\\
\sigma(\qq\,\rightarrow\,\cc) & = & \frac{8\pi}{27\sh}
\,\alpha_s^2(4m_c^2)\,\left (\sh+2mc^2\right )
\,\sqrt{1-\frac{4m_c^2}{\sh}}\,;\\
\\
& & \sh_{th}\ =\ 4\,m_c^2\,.
\end{array}
$$


\bigskip
$$
\begin{array}{lll}
\frac{d\sigma(gg\,\rightarrow\,\cc)}{d\th} & = &
\frac{\pi}{16\sh^2}\,\as^2(4m_c^2)
\,\left [\frac{12}{\sh^2}(m_c^2-\th)(m_c^2-\uh)
+\frac{8}{3}\frac{(m_c^2-\th)(m_c^2-\uh)-2m_c^2(m_c^2+\th)}
{(m_c^2-\th)^2}\right.\\
\\
& + &\frac{8}{3}\frac{(m_c^2-\th)(m_c^2-\uh)-2m_c^2(m_c^2+\uh)}
{(m_c^2-\uh)^2}\,-\,\frac{2m_c^2(\sh-4m_c^2)}
{3(m_c^2-\th)(m_c^2-\uh)}\\
\\
& - & 6\,\frac{(m_c^2-\th)(m_c^2-\uh)+m_c^2(\uh -\th)}
{\sh(m_c^2-\th)}
\left. \,-\,6\,\frac{(m_c^2-\th)(m_c^2-\uh)+m_c^2(\th -\uh)}
{\sh(m_c^2-\uh)}\right ]\,;\\
\\
\sigma(gg\,\rightarrow\,\cc) & = & \frac{\pi}{3\sh}
\,\as^2(4m_c^2)\,\left [ -\frac{1}{4}\left (7+\frac{31m_c^2}{\sh}
\right )\,x + \left (1+\frac{4m_c^2}{\sh}+\frac{m_c^4}{\sh^2}\right )
\,\log \frac{1+x}{1-x}\right ],\\
\\
 & & x\,=\,\sqrt{1-\frac{4m_c^2}{\sh}};\ \sh_{th}\ =\ 4\,m_c^2\,.
\end{array}
$$


\bigskip
$$
\begin{array}{lll}
\frac{d\sigma(qc\,\rightarrow\,qc)}{d\th} & = &
\frac{4\pi}{9(\sh-m_c^2)^2}\,\as^2(-\thmi)
\,\frac{(m_c^2-\uh)^2+(\sh-m_c^2)^2+2m_c^2\th}{\th^2}\,;\\
\\
\sigma(qc\,\rightarrow\,qc) & = & \frac{4\pi}{9(\sh -m_c^2)^2}
\,\as^2(-\thmi)
\,\left [ \left (1-\frac{2\sh}{\thmi}\right )
\left (\frac{(\sh-mc^2)^2}{\sh}+\thmi\right )
-2\sh\log\frac{(\sh-m_c^2)^2}{-\thmi\sh} \right ]\,;\\
\\
& & \thmi\ =\ 0;\ \ \thma\,=\,-(\sh-m_c^2)^2/\sh\,; \\
\\
& & \sh_{th}\ =\ m_c^2\,-\,\frac{1}{2}\,\thmi\,+\,
\left (-m_c^2\,\thmi\,+\,\frac{1}{4}\,\thmi^2\right )^{1/2}\,.
\end{array}
$$


\bigskip
$$
\begin{array}{lll}
\frac{d\sigma(gc\,\rightarrow\,gc)}{d\th} & = &
\frac{\pi}{16(\sh-m_c^2)^2}\,\as^2(-\thmi)\,
\left [\frac{32(\sh-m_c^2)(m_c^2-\uh)}{\th^2}\right.\\
\\
& + &\frac{64}{9}\frac{(\sh-m_c^2)(m_c^2-\uh)+2m_c^2(\sh +m_c^2)}
{\sh - m_c^2)^2}
\, +\, \frac{64}{9}\frac{(\sh-m_c^2)(m_c^2-\uh)+2m_c^2(\uh +m_c^2)}
{m_c^2-\uh)^2}\\
\\
& + &\frac{16}{9}\frac{m_c^2(4m_c^2-\th)}{(\sh-m_c^2)(m_c^2-\uh)}
\,+\,16\,\frac{(\sh-m_c^2)(m_c^2-\uh)+m_c^2(\sh-\uh)}
{\th(\sh-m_c^2)}\\
\\
& - &\left.\,\,16\,\frac{(\sh-m_c^2)(m_c^2-\uh)-m_c^2(\sh-\uh)}
{\th(m_c^2-\uh)}\right ]\,;\\
\\
\sigma(gc\,\rightarrow\,gc) & = & \frac{\pi}{(\sh-m_c^2)^2}
\,\as^2(-\thmi)
\,\left [ \left (1+\frac{4}{9}\left (\frac{\sh+m_c^2}{\sh-m_c^2}
\right )^2 \right )\,\left (\thmi-\thma\right )\right.\\
\\
& + & \frac{2}{9}\frac{\thmi^2-\thma^2}{\sh-m_c^2}
+2(\sh+m_c^2)\log \frac{\thmi}{\thma}\\
\\
& + & \frac{4}{9}\frac{\sh^2-6m_c^2\sh+6m_c^4}{\sh-m_c^2}
\log \frac{\sh-m_c^2+\thmi}{\sh-m_c^2+\thma}
\,+\,2\,(\sh-m_c^2)\left (\frac{1}{\thma}-\frac{1}{\thmi}
\right ) \\
\\
& + & \left. \frac{16}{9}\,m_c^4\,\left (\frac{1}{\sh-m_c^2+\thma}
\,-\,\frac{1}{\sh-m_c^2+\thmi}\right ) \right ]\,; \\
\\
\thma & = & -\max\left (\sh-m_c^2+\thmi,
(\sh-m_c^2)/\sh\right )\,,\\
\\
& & \left (\mbox{to enforce}\ \uh-m_c^2\,<\,\thmi\right )\,;\\
\\
\sh_{th} & = & \left\{
           \begin{array}{llll}
            m_c^2-\frac{1}{2}\thmi+\left ( -m_c^2\,\thmi
               \,+\,\frac{1}{4}\thmi^2\right )^{1/2}\,,
               & \mbox{if}\ -\thmi & < & \frac{1}{2}m_c^2\,;\\
               m_c^2\,-\,2\,\thmi\,,  & \mbox{if}\ -\thmi
               & > &\frac{1}{2}m_c^2\,.
           \end{array}
          \right.
\end{array}
$$
\nsz


\vfill\eject

\vfill\eject


\end{document}